\documentclass[aps,prl,reprint,twocolumn]{revtex4-1}

\usepackage{graphicx}
\usepackage{bm}
\usepackage{amsmath}
\usepackage{color}
\newcommand{\bee}{\begin{eqnarray}}
\newcommand{\eee}{\end{eqnarray}}

\begin{document}

\title{Periodic motion of sedimenting flexible knots}

\author{Magdalena Gruziel$^1$}
\author{Krishnan Thyagarajan$^2$}
\author{Giovanni Dietler$^2$}
\author{Andrzej Stasiak$^{3,4}$}
\author{Maria L. Ekiel-Je\.zewska$^1$}
\email{mekiel@ippt.pan.pl}
\author{Piotr Szymczak$^5$}
\email{piotr.szymczak@fuw.edu.pl}

\affiliation{$^1$Institute of Fundamental Technological Research, Polish Academy of Sciences, Pawi\'nskiego 5B, 02-106, Warsaw, Poland}
\affiliation{$^2$\'Ecole Polytechnique F\'ed\'erale de Lausanne, Lab Phys Living Matter, CH-1015 Lausanne, Switzerland}
\affiliation{$^3$Center for Integrative Genomics, University of Lausanne, CH-1015 Lausanne, Switzerland}
\affiliation{$^4$SIB Swiss Institute of Bioinformatics, CH-1015, Lausanne, Switzerland}
\affiliation{$^5$Institute of Theoretical Physics, Faculty of Physics, University of Warsaw, Pasteura 5, 02-093, Warsaw, Poland}

\date{\today}
\begin{abstract}
We study the dynamics of knotted deformable closed chains sedimenting in a viscous fluid. We show experimentally that trefoil and other torus knots often attain a remarkably regular horizontal toroidal structure while sedimenting, with a number of intertwined loops, oscillating periodically around each other. We then recover this motion numerically and find out that it is accompanied by a very slow rotation around the vertical symmetry axis. We analyze the dependence of the characteristic time scales on the chain flexibility and aspect ratio. It is observed in the experiments that this oscillating mode of the dynamics can spontaneously form even when starting from a qualitatively different initial configuration. In numerical simulations, the oscillating modes are usually present as transients or final stages of the evolution, depending on chain aspect ratio and flexibility, and the number of loops. 

\end{abstract}
\maketitle

\indent Long and flexible strings tend to get knotted~\cite{Belmonte2001,bennaim2001,Belmonte2007,Raymer2007}, a fact known to anyone who has taken earphones out of the pocket only to find them hopelessly tangled. The same is true also at the microscale - one finds knots in polymer chains~\cite{micheletti2011}, DNA~\cite{Bates1993} and proteins~\cite{Virnau2006}. Nevertheless, the relation between the topological constraints and dynamics and function of biomolecules remains elusive~\cite{Meluzzi2010}. \\
\indent There are many examples when the presence of a knot affects {\it local} properties of the polymer chain. Knotting was shown to locally weaken the chain, so that it breaks when pulled at the site of the knot, no matter if we deal with polymer chains \cite{saitta1999}, ropes \cite{audoly2007} or spaghetti \cite{Pieranski2001}. In a similar vein, tight knots on DNA and proteins increase the local width of the chain, which can lead to jamming of the nanopores \cite{Rosa2012} and mitochondrial pores \cite{Szymczak2016}, when the pulling force is sufficiently high. \\
\indent Much more subtle are the connections between the topological constraints and the {\it global} properties of the chain. In this respect, it has been hypothesized that knots in proteins provide extra stability needed to maintain the global fold and function under harsh conditions~\cite{Taylor2007}. Indeed, both experiments \cite{Sayre2011} as well as numerical simulations \cite{Sulkowska2008} suggest that knotted structures have an increased thermal stability when compared with unknotted ones, which can potentially explain why the knotted structures are encountered in thermophilic bacteria \cite{Nureki2002}. Another manifestation of the impact of topology on the global properties of the chain is the correlation between the knot type of the DNA loops and their electrophoretic mobility \cite{stasiak96,michieletto2015}. The latter turns out to be linearly dependent on the average crossing number of the knot. Thus, electrophoresis can be used for the  separation of DNA topoisomers. A similar linear relationship holds for sedimentation of DNA knots~\cite{volog98}, as well as macroscopic rigid knots \cite{weber13}. \\
\indent In this Letter, we investigate shapes and dynamics of sedimenting flexible knotted loops. In our experiments, knotted chains made of millimeter-sized steel balls settle in a viscous oil, and might also serve as a model of microscale filaments moving in a water-based environment. Surprisingly, we observe that trefoil and other torus knots often attain remarkably regular, thin, wide and relatively flat horizontal toroidal structures, with a number of intertwined loops which perform an oscillatory periodic motion. This characteristic motion is often visible even if starting from different initial configurations. Interestingly, a similar motion has been reported previously in a completely different context of knotted vortices in ideal fluid \cite{keener90,ricca99,ricca93}.  \\
\indent We then study this motion in detail using Stokesian Dynamics simulations of a flexible bead-spring chain moving under gravity. For torus knots, we numerically find translating periodic solutions analogous to those seen in the experiments, and we detect that they slowly rotate along the vertical symmetry axis. We analyze the dependence of the characteristic time scales on the chain flexibility and aspect ratio. \\
\indent The results are important for elucidating the link between topology and dynamics  of filamentous objects, necessary e.g. for a correct interpretation of ultracentrifugation or sedimentation data~\cite{Liu1981,Roovers1983,Rybenkov1997}. The usual assumption adopted in interpretation of the ultracentrifugation data is that the polymer settles in a globular form, whereas our studies 
of relatively short closed chains 
show that it does not need to be so. Torus-like forms have different sedimentation velocities from the globules made of the same length of polymer, which needs to be taken into account while interpreting the ultracentrifugation data.

Moreover, the existence of periodic solutions for flexible knotted chains is essential for such general features of the dynamics as bifurcations, instabilities and transition to chaos, which can be further analyzed in analogy to previous studies performed for simple model systems \cite{janosi97,felderhof2005,jung06,mej14,gruca15} of many particles settling under gravity in a viscous fluid, with a striking coexistence of both periodic oscillations and chaotic trajectories \cite{janosi97,gruca15}.

The present study is an important extension of previous 
 investigatons of flexible open filaments settling in a gravitational field \cite{llopis2007,li2013,saggiorato2015,bukowicki15} and of knotted filaments in the shear flow~\cite{Matthews2010,Kuei2015,liebetreu2018}, which show that
an interplay between elastic and hydrodynamic forces can lead to a remarkably rich dynamical behaviour.

\begin{figure}[b]\vspace{-0.3cm}
\includegraphics[width=0.49\textwidth]{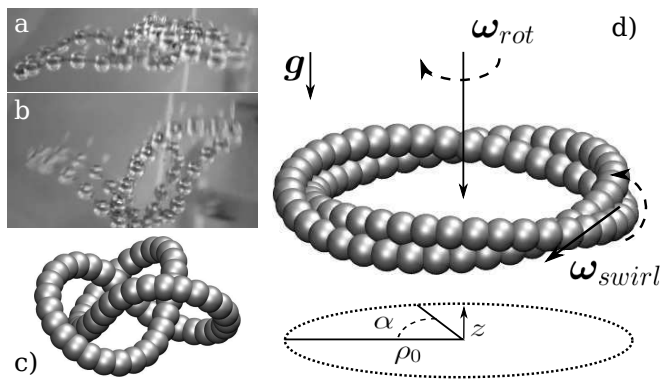}\vspace{-0.1cm}
\caption{\label{fig:init} Examples of structures evolving to regular configurations  $\mathcal{T}_{2,3}$; a,b) experiments; c) simulations; d) the notation used in this article, with $\omega_{rot}\!=\!2\pi/\tau_{rot}$ and $\omega_{swirl}\!=\!2\pi/\tau_{swirl}$. In all images gravity points downwards. }
\end{figure}

\begin{figure}[h]\vspace{-0.2cm}
\includegraphics[width=0.49\textwidth]{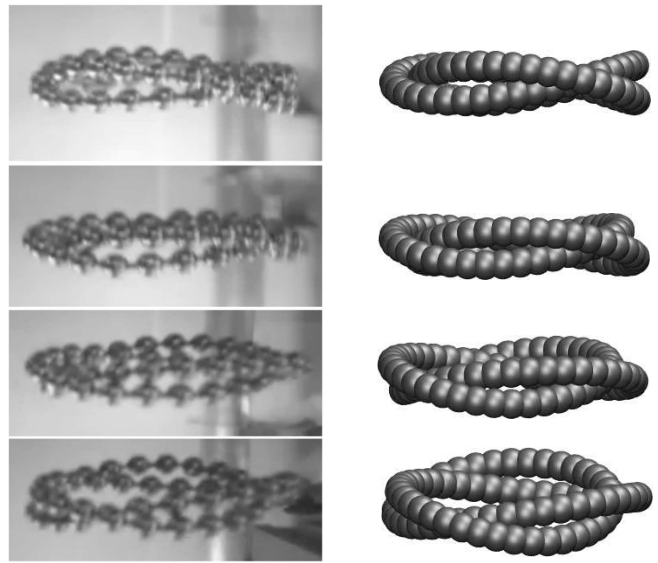}
\vspace{-0.6cm}
\caption{\label{fig:snaps} Snapshots of sedimenting $\mathcal{T}_{2,3}$ knots from experiments (left) and simulations (right).
}
\end{figure}

%
\noindent{\bf Experiments.} Chains made of 50 stainless steel spherical beads of diameter 4.5mm connected with 2mm linkers were joined end-to-end using a stainless-steel clip, in such a way that knotted loops were formed. Knots of different types  were studied, including $3_1,\,4_1,\,5_1,\,5_2,\,6_1,\,7_1$.  The experiments were conducted in highly viscous (1000cSt) silicone oil that filled the cylinder 2m high and 0.5m in diameter. The chains were dropped into oil, with the total number of 350 trials and typical Reynolds number  of the order of 1-10 (based on characteristic radius of the sedimenting structure). The mean sedimentation velocity $V$ was measured and videos of the sedimenting chains were recorded. \\
\indent We observed different behaviors of the sedimenting knots, depending on the knot type and initial configuration (see an example in Fig.~\ref{fig:init}a,b). 
In many trials the chains tended to regular or almost regular sequences of shapes, although in some cases they settled as a compact coiled structure, sometimes staying in the same initial configuration until the end of the experiment. The torus knots $3_1,\,5_1,\,7_1,\,9_1$ had a tendency to converge onto a horizontal, flat and wide toroidal shape with a circular hole at the center and a number of tightly intertwined loops, which swirled around each other  periodically (or almost periodically). This characteristic attracting configuration is in this Letter  called $\mathcal{T}_{p,q}$, with the indices referring to  $p$ loops, each making $q$ turns around the centerline of the torus.  \\
\indent The simplest torus knot, trefoil ($3_1$), often evolved towards oscillating shape $\mathcal{T}_{2,3}$ (see Fig.~\ref{fig:snaps} and movie 1)
, with the radius $\rho_0$=2.6 cm and performed about 5 swirling oscillations over the 1.6m distance of the observed motion \footnote{The first and last 0.2m of the sedimentation path were discarded from the analysis to avoid the impact of the top and bottom boundaries and to allow the structures to relax their configurations.}. The experimental data allows to estimate that for the $\mathcal{T}_{2,3}$ shape, the typical ratio of the swirling and sedimentation time scales, $\tau_{swirl}$ and $\tau_{sed}\equiv 2\rho_0/V$, is of the order of 6. To demonstrate the swirling motion, we colored one of the beads in a video recording, and added it as Movie 7 to the SM~\footnote{See Supplemental Material at [URL will be inserted by
publisher], which includes Ref.~\cite{allen}, for additional
details of the simulations}.
For certain initial conditions of a trefoil, another configuration $\mathcal{T}_{3,2}$ was formed and swirling motion appeared (see  movie 2 and Figs.~1-2 in SM). 

\indent More complicated torus knots studied in the experiments in many cases also formed analogous toroidal structures, e.g. $\mathcal{T}_{2,7}$ (see movie 3) or $\mathcal{T}_{2,5}$, with a corresponding periodic motion of the loops around each other. However, the experimental data do not allow for a thorough characterization of the periodic orbits. This has motivated us to search for such dynamical modes of torus knots numerically, as detailed below.\\
%
\noindent{\bf Theoretical model.} Elastic loop is modeled as a chain of N spherical beads of diameter $d$ connected by harmonic springs. The stretching potential is thus given by $E_{b}=\sum_i B (l_i-l_0)^2/2$, where $l_i$ is the distance between centers of beads $i$ and $i+1$. 
The equilibrium distance is chosen to be $l_0=0.6d$ (the beads overlap to prevent fluid motion between them~\cite{slowicka15}). With this choice, $N$=120 corresponds to the same aspect ratio as for the chains used in the experiments (for completeness, we analyze numerically also motion of shorter loops). The spring constant is set at $B=50 F_0/d$, where $F_0$ is a gravitational force acting on each of the beads. A relatively high value of $B$ assures that the length of the chain is almost constant in time. \\
\indent We introduce also a harmonic bending potential, $E_{a}\nobreak=\nobreak\sum_i A (\phi_i-\phi_0)^2/2$, where $\phi_i$ is the angle between the bonds $(i-1,i)$ and $(i,i+1)$. A moderate bending stiffness $A\!=\!5 F_0 d$ is assumed, whereas the chosen equilibrium value of the angle, $\phi_0\!=\!\pi$, corresponds to a straight chain. Additionally, a truncated Lennard-Jones potential $E_{rep} = \varepsilon((\sigma/r_{ij})^{12}-(\sigma/r_{ij})^{6}), \  r_{ij} < 2^{1/6}\sigma$ is used to represent steric constraints between the non-consecutive beads, with $\sigma=0.5d$ and $\varepsilon=0.25 d F_0$. \\
\indent To follow the dynamics of the system, one needs to find the fluid flow produced by the sedimenting chain. For macroscopic objects, this task is hard in general due to the nonlinearity of the Navier-Stokes equations. However, our ultimate goal is to understand sedimentation process of microscopic knotted loops. Therefore, to make the problem tractable, and applicable to microscale systems, we assume that the motion takes place at very low Reynolds number. The inertial term in the Navier-Stokes equations can then be neglected and the relation between the bead velocities and the forces acting on them becomes linear
\begin{equation}
\dot{\mathbf{r}}_i = \sum_{j=1}^N \bm{\mu}_{ij}\cdot\mathbf{F}_j,
\end{equation}
\noindent where $\mathbf{r}_i$ is  the position of the center of bead $i$, $\bm{\mu}_{ij}$ are translational-translational mobility matrices and $\mathbf{F}_i$ is the total external force acting on bead $i$ (i.e. the sum of the gravitational force $F_0$ and the forces resulting from the interparticle potentials described above). \\ 
\indent For the mobility, we adopt the Rotne-Prager approximation~\cite{rp69,yamakawa70,wajnryb13}, which for $i\! \neq \!j$ gives $\bm{\mu}_{ij}/\mu_0=\frac{3d}{8r_{ij}}((1+\frac{d^2}{6r_{ij}^2})\bm{I}+(1-\frac{d^2}{2r_{ij}^2})\hat{\bm{r}}_{ij}\hat{\bm{r}}_{ij})$ for the nonoverlapping beads ($r_{ij}\equiv|\mathbf{r}_i-\mathbf{r}_j|\ge d$) and  $\bm{\mu}_{ij}/\mu_0=(1-\frac{9r_{ij}}{16d})\bm{I}+\frac{3r_{ij}}{16d}\hat{\bm{r}}_{ij}\hat{\bm{r}}_{ij}$ for the overlapping beads ($r_{ij}<d$). Finally, the self term is given by $\bm{\mu}_{ii}\!=\!\mu_0\bm{I}$. Here, $\bm{I}$ is the 3x3 unit matrix and $\mu_0\!=\!(3 \pi \eta d)^{-1} $ is the single bead mobility coefficient. \\ 
%
\noindent{\bf Numerical results and discussion.} In our simulations, we focused on evolution of torus knots, starting from toroidal structures with the symmetry axis parallel to gravity. Some of the initial configurations (an example is shown in Fig.~\ref{fig:init}c) were chosen on surfaces of horizontal tori with different values of the major and minor radii, $r_0$ and $a$, respectively, with the bead centers located along the line which in cylindrical coordinates is given by
\begin{equation}
\rho(\alpha) \! =\! r_0 + a \sin\left(\frac{q}{p} \alpha\right), \ \ \ \ z(\alpha)  \!= \!a \cos\left(\frac{q}{p} \alpha\right)
\label{eq:tknot}
\end{equation}
with the angle $\alpha \in (0,2 p \pi)$, the number of loops $p$, and the number of turns in poloidal direction $q$.\\
\indent The initial structures given by Eq.~\eqref{eq:tknot} later often  increased their major radius up to $\rho_0\! \!\approx \!\!Nl_0/(2\pi p)$, decreased the minor radius, deformed the minor cross-section from circular to elliptic, and acquired the swirling ${\cal T}_{p,q}$ shape, akin  to that observed in the experiments (see movies 4-6 in SM), and lasting for a shorter or longer time, depending on the knot type, fiber length $N$ and bending stiffness $A$. In the following, we focus on analyzing the longest and the most regular swirling motion of the $\mathcal{T}_{2,3}$ structure. Snapshots from the simulations are presented in Fig.\,\ref{fig:snaps} along those from the measurements, clearly showing the similarity of the experimental and numerical shapes, also visible by comparing movies~1 and~4. \\
\indent In our simulations, we were able to follow the sedimentation process over a much longer time than in the experiments - such that the  distance travelled by a torus knot ${\cal T}_{p,q}$ of radius $\rho_0$ was at least $\sim\!\! 2\!\cdot\! 10^3 \rho_0\!$. Such a long time frame allowed us to observe several features of the dynamics which are hard to perceive in the experiments. In particular, we note that the swirling motion of the strands around each other (shown in Fig.\,\ref{fig:snaps}) is accompanied by a much slower rotation of the entire system around the vertical symmetry axis. The direction of rotation depends on the handedness of the knot. 
The simulation shows that the  right-handed trefoil rotates clockwise and left-handed -- anti-clockwise, looking from the top. (A knot called right-handed if its writhe is positive and left-handed otherwise \cite{liang1994}). This stays in accordance with experimental results reported for rigid knots~\cite{weber13}.

On the other hand, the direction of the swirling motion of the strands is always the  same, independent of chirality of the knot. The outer strand (i.e. the one positioned further from the center of the torus) lags behind the inner one. Then, the strands swap places and the process repeats itself. This dynamics can be understood by noting that the beads in the inner strand are on average closer to all the other beads than the beads in the outer strand. Closer beads exhibit stronger hydrodynamic interactions, which results in faster sedimentation and overtaking of the slower strand. Actually, the same mechanism of hydrodynamic interactions induced by gravity is responsible for swirling motions of many rigid particle systems settling under gravity \cite{janosi97,felderhof2005,jung06,mej14,gruca15}, and also swirling motions of two elastic filaments \cite{llopis2007}, and two elastic dumbbells~\cite{bukowicki15}.

\indent Thus, in our simulations, there emerge  three characteristic time scales: period $\tau_{rot}$ of the system rotation, period $\tau_{swirl}$ of the swirling oscillations, and sedimentation time  $\tau_{sed}=2\rho_0/V$, i.e.
the time needed for a chain to sediment over its diameter.  The corresponding frequencies, are determined from the Fourier transform of the bead positions, and the sedimentation velocity $V$ by averaging the center-of-mass speed. \\
\begin{figure}[h]\vspace{-0.5cm}
\includegraphics[width=0.99\columnwidth]{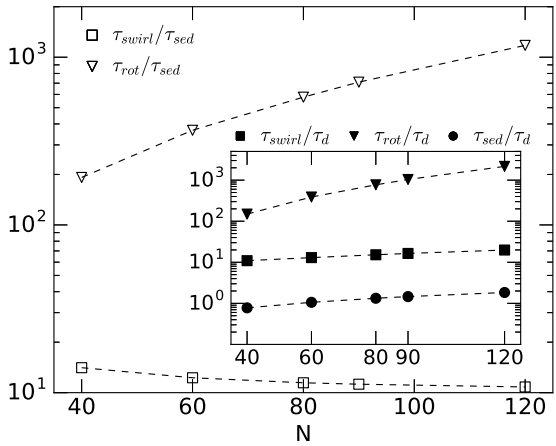}
\caption{\label{fig:freqs} Ratios of timescales for swirling, rotation and sedimentation of $\mathcal{T}_{2,3}$ loops vs their length (numerical results). 
}
\end{figure}
\indent Ratios of the characteristic timescales for $\mathcal{T}_{2,3}$ are plotted in Fig.~\ref{fig:freqs}. Here, $\tau_d=d/(\mu_0 F_0)$ is the time needed by a single bead falling under the gravitational force $F_0$ to move over the distance equal to its diameter $d$. \\
\indent The essential observation is the separation of timescales, $\tau_{sed} \!\ll \! \tau_{swirl}\! \ll \!\tau_{rot}$ in the simulations. The simulation data in Fig.~\ref{fig:freqs} give the numerical ratio $\tau_{swirl}/\tau_{sed}\!\approx \!10$ -- of the same order of magnitude as in experiments. On the other hand, a very long period of rotation around vertical axis   makes the rotation difficult to be  observed in experiments. Moreover, the separation of timescales still holds, even if the bending stiffness $A$ is changed by 4 orders of magnitude (see Fig.11 in SM).\\
\indent
Flat, thin and wide toroidal shape $\mathcal{T}_{2,3}$ closely resembles an equilibrium configuration of an elastic trefoil made of springy wire, as reported in Refs. \cite{buck2004,gallotti2007,starostin2014,Gerlach2017}, without gravity or fluid involved. These configurations are found by minimizing the bending energy of the knotted loop. 
Such flat toroidal shapes $\mathcal{T}_{p,q}$ will be equilibrium configurations for all the knots with the bridge index equal to the braid index~\cite{gallotti2007}, which is a fairly large class of knotted structures, including (but not limited to) torus knots. On the other hand, the swirling motion of the strands around each other is very similar to the toroidal vortex knot solutions in inviscid fluid \cite{keener90,ricca99}. The latter, for large aspect ratio of the chain, lie on an elliptic torus of eccentricity $q/\sqrt{(p^2+q^2)}$ and swirl around each other periodically. This has motivated us to ask whether the $\mathcal{T}_{2,3}$ solutions of the sedimenting knots can be described by the same mathematical expressions as in case of vortex knots. Therefore, we check if all the beads move along the same trajectory
 \bee
 \rho(\alpha) &\!=\!& \rho_0  + \rho_0 \epsilon \sin \left(\frac{q}{p} \alpha\right)\!,\hspace{0.3cm}
 z(\alpha)\!=\! \rho_0 \epsilon \frac{\sqrt{p^2\!+\!q^2}}{q}\cos \left(\frac{q}{p} \alpha\right)\!,\nonumber\\ \label{eq:structure}
 \eee
\noindent on the surface of an elliptic torus, and if their cylindrical coordinates change in time according to the analogous formula~\cite{ricca99,ricca93},
\begin{equation}
\left\{
\begin{aligned}
\rho(t) & = \rho_0 + C \sin\left(\frac{2 \pi t}{\tau_{swirl}}\right), \\
\alpha(t) & = C \frac{p}{q\rho_0} \cos\left(\frac{2 \pi t}{\tau_{swirl}}\right) + \frac{2\pi t}{\tau_{rot}},\\
z(t) & = C \frac{\sqrt{(p^2\!+\!q^2)}}{q} \cos\left(\frac{2 \pi t}{\tau_{swirl}}\right) + Vt.
\end{aligned} 
\right. \label{eq:fits}
\end{equation}
\noindent where $\rho$, $z$, $\alpha$ should be referred to discrete positions of consecutive bead centers (we focus on $p\!\!=\!\!2$ loops and $q\!\!=\!\!3$ turns). Here, $\rho_0$ corresponds to the major radius of the torus, and $C\!\!\ll\!\! \rho_0$ is the shorter radius of elliptic cross-section. It is worth emphasizing here, that the Eq.~\eqref{eq:fits} includes three contributions to the overall motion: swirling of the loops with a period $\tau_{swirl}$, sedimentation with a velocity $V$, and rotation of the system around vertical axis with a period $\tau_{rot}$. Both $\rho_0$ and $C$ are evaluated from fitting to the simulation data. The results are shown in Fig.~\ref{fig:fits} for the chain aspect ratio $N$=90 and in the SM for $N$=40 and 120. One should note, that for the case of $\mathcal{T}_{2,3}$ knot and the sizes considered here, fitted $\rho_0$ is smaller by less than 1\% from the simple geometric estimate, $\rho_0=Nl_0/(2\pi p)$. The scaled coordinates used in these Figures correspond to \mbox{$\tilde{\rho}(t)=C^{-1}(\rho(t)-\rho_0)$}, \mbox{$\tilde{\alpha}(t)=\frac{q\rho_0}{p}C^{-1}(\alpha(t)-\frac{2\pi t}{\tau_{rot}})$}, \mbox{$\tilde{z}(t)=\frac{q}{\sqrt{p^2+q^2}}C^{-1}(z(t)-Vt)$}, and the time coordinate is scaled by the swirling period of the chain, $\tau_{swirl}(N)$. It is evident that the fitting works rather well. \\
\begin{figure}[b]\vspace{-0.3cm}
\includegraphics[width=0.49\textwidth]{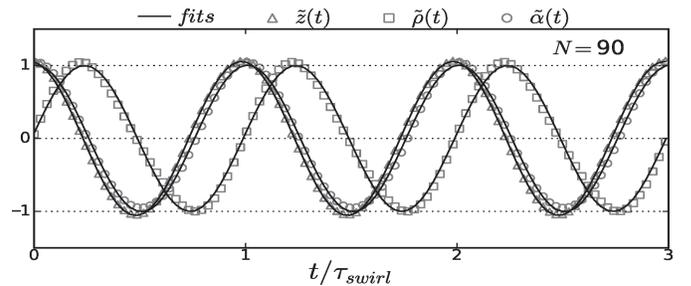} \vspace{-0.4cm}
\caption{\label{fig:fits} Swirling motion of a single bead of the ${\cal T}_{2,3}$ chain with $N=90$, in scaled cylindrical coordinates $\tilde{\rho},\tilde{\alpha},\tilde{z}$ (see text) represented with squares, circles, and triangles, respectively. Solid lines are fits to the simulation data (see SM).} 
\end{figure}
\noindent For $40\!\!\le\!\! N\!\le\!\! 120$, we obtain practically the same value $C/d\!\!=\!\!0.28\pm 0.01$. For this range of aspect ratios, $1.91 \!\le \!\rho_0/d\!\le\! 5.73$ and therefore $C/\rho_0\!\!\ll\!\! 1$, consistently with the range of validity of the expressions for vortex knots \cite{ricca99,ricca93}.  \\ 
\indent Figs.~\ref{fig:init}-\ref{fig:fits} present the data for $\mathcal{T}_{2,3}$ structure. A natural question to ask is whether the motion of other toroidal structures, with different $p$ and $q$, exhibit similar features. Several of them have been studied within this work and some of the results are given in SM. Briefly, the $\mathcal{T}_{3,2}$ motion is almost periodic with a reasonably well defined swirling frequency and very irregular amplitudes, see Fig.~2 in SM. On the other hand, the $\mathcal{T}_{2,5}$, $\mathcal{T}_{2,7}$, and $\mathcal{T}_{2,9}$ knots yield regular, periodic motion as in the case of $\mathcal{T}_{2,3}$ (see Figs.~4-6 in SM), though are less stable, in the sense of the distance travelled. 
The above mentioned stability, and the applicability of Eq.~\eqref{eq:fits} would also depend on the size $N$ or bending stiffness, $A$, of the chain.  \\
\indent The  particles oscillating around nearly circular poloidal orbits have also been observed (numerically and analytically) for systems of two rings of particles without elastic constraints, sedimenting very close to each other under a perpendicular constant force \cite{mej14}. The poloidal swirling of particles in a toroidal configuration is the inherent feature of hydrodynamic interactions generated by the external force parallel to the symmetry axis of the torus. Knotted flexible chains can keep the toroidal ${\cal T}_{p,q}$ configuration for a long time, and therefore for such systems the swirling motion might be significant for practical applications. Importantly, even in the presence of fair amount of Brownian motion (characterized by the Peclet number Pe=$\frac{dF_0}{kT}\gtrsim 2$) both the swirling motion persits and the toroidal shape is still present, although from time to time it flips over due to the thermal noise (see movie 8 with Pe=2 in SM). 
\\
%
\indent In summary, our experiments have shown that knotted loops made of ball chain, when sedimenting, can attain a flat, wide and thin toroidal form, with a number of intertwined loops oriented perpendicularly to gravity. Such a structure moves in a highly coordinated fashion, with the individual loops swirling periodically around each other. In numerical simulations of Stokesian dynamics of elastic fibers, a family of oscillating motions has been found, with a striking similarity to those seen in the experiments. 
 Additionally, the periodic solutions obtained in the simulations resemble the stable solutions of knotted vortex filaments evolution equations in an ideal fluid. \\ 

\begin{acknowledgments} 
This work was supported in part by Narodowe Centrum Nauki under grant 2015/19/D/ST8/03199. M.G. and M.L.E.-J. benefited from the COST Action MP1305. We thank Renzo Ricca and Tony Ladd for helpful discussions. Figures and movies visualizing the results of simulations were prepared with help of VMD package~\cite{Humphrey1996}.
\end{acknowledgments}

\end{document}